\documentclass[12pt]{iopart}
\usepackage{graphicx}
\begin{document}

\title[Sub-wavelength imaging in the canalization and resonant tunnelling regimes]{Comparison of imaging with sub-wavelength resolution in the canalization and resonant tunnelling regimes}

\author{R.  Koty{\'n}ski and T. Stefaniuk}

\address{Department of Physics, University of Warsaw, Pasteura 7, 02-093 Warsaw, Poland}
\ead{rafalk@fuw.edu.pl}
\begin{abstract}
We compare the properties of subwavelength imaging in the visible wavelength range for metal-dielectric multilayers operating in the canalization and the resonant tunnelling regimes. The analysis is based on the transfer matrix method and time domain simulations. We show that Point Spread Functions for the first two resonances in the canalization regime are approximately Gaussian in shape. Material losses suppress transmission for higher resonances, regularise the PSF but do not compromise the resolution.
In the resonant tunnelling regime, the MTF may dramatically vary in their phase dependence. Resulting PSF may have a sub-wavelength thickness as well as may be broad with multiple maxima and a rapid phase modulation. We show that the width of PSF may be reduced by further propagation in free space, and we provide arguments to explain this surprising observation.
\end{abstract}

\vspace{2pc}

\maketitle

\section{Introduction}
In 2000, Pendry~\cite{pendry2000nrm} has shown that a $40nm$ silver slab is capable of  near filed imaging with resolution beyond the diffraction limit in the near UV wavelength range. 
The operation of a similar but asymmetric~\cite{anantharamakrishna2002aln} flat lens has been later experimentally verified~\cite{melville2005sri,fang2005sdl}. In these works, the mechanism leading to  superresolution is related to the appearance of surface plasmon polariton (SPP) mode, which enables the transfer of evanescent part of the spatial spectrum through the slab. Other ways of achieving superlensing include the use of photonic bandgap materials showing negative refraction~\cite{notomi2000tlp}. The negative refraction is similar as in left-handed metamaterials discussed forty years ago by Veselago~\cite{Veselago68} who studied theoretically the properties of materials with a negative real part of both permittivity and permeability.

\begin{figure}
\begin{center}
a.\includegraphics[width=7cm]{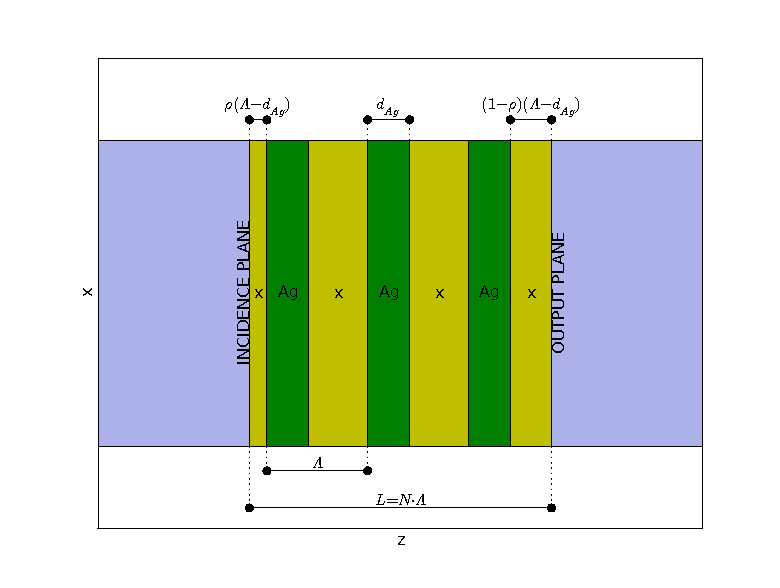}
b.\includegraphics[width=7cm]{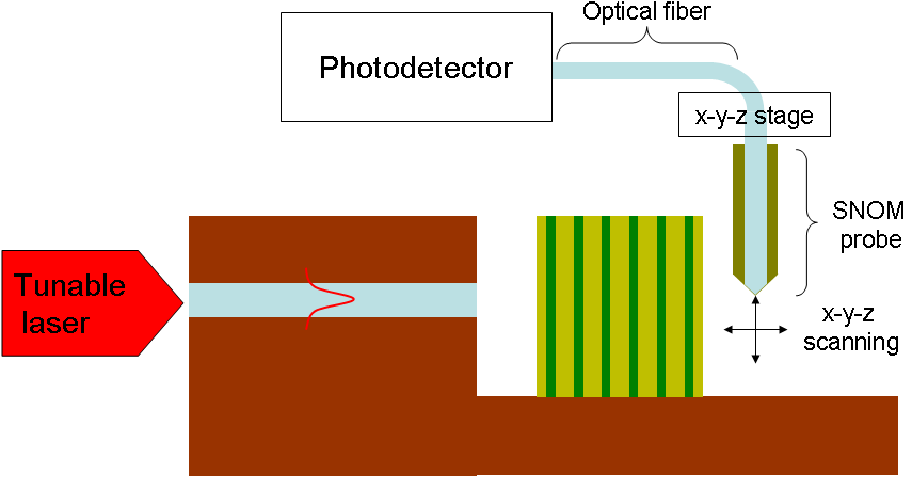}
\end{center}
\caption{a. Schematic view of the multilayer consisting of $N$ periods of thickness $\Lambda$. Altogether the multilayer contains $2N+1$ layers of silver, or dielectric $x$; \textbf{b.} Schematic view of an experimental set-up for the characterisation of imaging through the multilayer~\cite{Fabre:PRL-101-073901}.}
\label{schematicview}
\end{figure}
 
Moreover, also the evanescent waves are subject to amplitude enhancement in the left-handed materials. Therefore, the imaging through a slab with a Veselago material may be, at least theoretically, truly lossless and aberrationless.
For the Pendry's superlens, both losses as well as the cut-off wavelength of the SPP mode limit the operation range to subwavelength distances only~\cite{webb2004mnr,smith2002lsd}. This obstacle has been tackled to some degree in several ways. One of them is by breaking the silver slab into a metallic multilayer~\cite{anantharamakrishna2002aln,wood2006} with effective anisotropic properties and strong coupling of the SPP modes between neighbouring layers. Otherwise the loss may be compensated by the use of media with gain~\cite{anantharamakrishna2003raa,ponizovskaya2007mni}. Unfortunately, SPP modes introduce singularities into the modulation transfer function (MTF)~\cite{Saleh,GoodmanFourierOptics}, resulting in deteriorated imaging~\cite{anantharamakrishna2002aln,dorofeenko2006fwa,nietovesperinas2004pis,melville2006josab}.

The aim of this paper is to characterise the sub-wavelength near-field imaging properties of superlenses operating in two other, recently discussed regimes. 
These operation regimes are the resonant tunnelling~\cite{Scalora2007opex,Scalora1998ap} and canalization~\cite{belov2006prb,belov2005csi,ikonen2006eds}. Unlike the previously discussed regimes, which used SPP or negative refraction, resonant tunnelling and canalization rather depend on the band structure of the metal-dielectric one-dimensional photonic crystal, its effective properties, and the coupling properties of the external layers. We analyse structures with a high transmission, with thicknesses ranging up to several wavelengths - working under resonant tunnelling, or thinner than wavelength but with very regular responses - working under canalization. We also address the role of losses and finite layer thickness for the canalization regime. In our analysis we characterise these superlenses with the point spread function (PSF)~\cite{Saleh,GoodmanFourierOptics}, and their FWHM and standard deviations. 
The potential applications of these elements are similar to other superlenses and include eg. nanolithography~\cite{fang2005sdl}, and proximity printing~\cite{melville2006josab}.

\section{Background}
\subsection{Numerical method}
Our numerical analysis combines two elements: 1. the transfer matrix method (TMM)\cite{BornWolf}, which we use to examine the transmission of a single plane-wave through the multilayer, and 2. the theory of linear shift-invariant systems (LSI)~\cite{Saleh,GoodmanFourierOptics}, which provides the framework for the description of the multilayer with the modulation transfer function (MTF) or the point spread functions (PSF). In this section we review the principles of both elements. Appart from TMM, we also include finite difference time domain (FDTD)~\cite{Taflove:FDTD} simulations which are performed using a freely available software package MEEP with subpixel smoothing for increased accuracy~\cite{Farjadpour2006ol}.

Linear shift invariant scalar systems (LSI) are well known in different fields of Optics. Here we apply this model for the description of optical multilayers in a situation when they act as imaging elements for coherent monochromatic light.

Let us introduce some denotations that are used throughout the entire text. Figure~\ref{schematicview}a shows a schematic view of a metal-dielectric periodic multilayer. The structure is surrounded by air, and consists of periodically arranged metallic ($Ag$) and dielectric layers (made of a dielectric denoted as $x$). The structure has the total thickness $L$, contains $N$ periods with the pitch $\Lambda$, and single layer thicknesses of $d_{Ag}$ and $d_x$, respectively. First and last layers have the thickness dependent on parameter $\rho$. The initial layer of $x$ has the thickness of $\rho (\Lambda-d_{Ag})$, and the last has the thickness of $(1-\rho)(\Lambda-d_{Ag})$. The incident light is coherent and monochromatic with the free-space wavenumber $k_0=2\pi/\lambda$. The time dependence of the complex fields is $exp(-\imath k_0 c t)$.

In-plane imaging through a layered structure consisting of uniform and isotropic materials represents a LSI, for either TE or TM polarisations. Linearity of the system is the consequence of linearity of materials and validity of the superposition principle for the optical fields. Shift invariance results from the assumed infinite perpendicular size of the multilayer and the freedom in the choice of an optical axis. A scalar description is valid for the TE and TM polarisations in 2D since all other field components may be derived from $E_y$ or $H_y$, respectively.
For instance, for the TM polarisation, the magnetic field $H_y(x,z)$ may be represented with its spatial spectrum $\hat H_y(k_x,z)$
\begin{equation}
 H_y(x,z)=\int_{-\infty}^{+\infty} \hat H_y(k_x,z) exp(\imath k_x x) dk_x,
\end{equation} 
where, at least for lossless materials, the spatial spectrum is clearly separated into the propagating part $k_x^2<k_0^2 \epsilon$ and evanescent part $k_x^2>k_0^2 \epsilon$.

LSI can be characterised either using the modulation transfer function (MTF), or equivalently with the point spread function (PSF)~\cite{Saleh,GoodmanFourierOptics}. The MTF is simply the ratio of the spatial spectra of the output and incident fields and therefore corresponds to the amplitude transmission coefficient of the multilayer
\begin{equation}
 \hat H_y(k_x,z=L)=MTF(k_x) \cdot \hat H_y^{Inc}(k_x,z=0).
\end{equation}
We note that due to reflections, the incident field $\hat H_y^{Inc}(k_x,z=0)$ differs from the total field $\hat H_y(k_x,z=0)$.

MTF is obtained with the TMM method. For this purpose we analyse separately the transmission of plane-waves with different real values of the wavevector component $k_x$, corresponding to both propagating and evanescent waves. In each layer, the 
wavevector component $k_x$ is conserved, and $k_z^l=\pm\sqrt{k_0^2\epsilon^l-k_x^2}$. The magnetic field in layer $l$ may be written as
\begin{equation}
 H_y^l(x,z)=A_{+}^{l}exp(\imath(k_x x+k_z^l z))+A_{-}^{l}exp(\imath(k_x x-k_z^l z)),\label{eq.ampl}
\end{equation}
where the amplitudes $A_{+}^l$ and $A_{-}^l$ are matched at layer boundaries to satisfy the continuity of both $H_y$ and $E_x$. This leads to the transfer matrix relation between subsequent layers
\begin{equation}
 \left[\begin{array}{c} A_{+}^{l} \\ A_{-}^{l}\end{array}\right] = \left[ \begin{array}{cc} e^{-\imath z_l k_z^l} & 0 \\ 0 & e^{\imath z_l k_z^l} \end{array} \right]  \cdot  \left[ \begin{array}{cc} 1 \over{1-r} & -r \over{1-r} \\ -r \over{1-r} & 1 \over{1-r} \end{array} \right]  \cdot \left[ \begin{array}{cc} e^{\imath z_l k_z^{l+1}} & 0 \\ 0 & e^{-\imath z_l k_z^{l+1}}  \end{array} \right]  \cdot  \left[ \begin{array}{c} A_{+}^{l+1} \\ A_{-}^{l+1}\end{array}\right], \label{eq.tm}
\end{equation}
where $z_l$ is the position of the boundary, and $r$ is the Fresnel reflection coefficient for $H_y$ at the boundary. Solving the linear system of equations from eq.~(\ref{eq.tm}) gives the complex transmission and reflection coefficients of the multilayer, and together with eq.~(\ref{eq.ampl}) allows to reproduce the field inside the structure.
We will publish separately the details of the numerical method which fall beyond the scope of this paper. Let us only note that the 1. TMM further simplifies for periodic structures, 2. using a moving coordinate system for the amplitudes $A_{-}^{l}$ and $A_{+}^{l}$ improves stability of the TMM for evanescent waves, 3. a full 3D analysis is more involved\cite{Kotynski2008} and falls beyond the scalar model, 4. the present description is compatible with plane-wave waveguide solvers, eg.\cite{Kotynski2007oqe}, which may easily serve as the source field generator.

PSF is the inverse Fourier Transform of the MTF and has the interpretation of the response of the system to a point signal $\delta(x)$. The response to an arbitrary input $H_y^{Inc}(x,z=0)$ can be further expressed as its convolution with the PSF
\begin{equation}
 H_y(x,z=L)=H_y^{Inc}(x,z=0)* PSF(x)
\label{eq.psfconool}
\end{equation} 

While the MTF and PSF represent the imaging system in an equivalent way, and MTF is easier to calculate, there exist obvious advantages of using PSF. PSF can be more straightforwardly interpreted, as it provides visible information about the resolution, loss or enhancement of contrast, as well as the character of image distortions. We use the standard deviation and full-width-at-half-maximum (FWHM) to compare the characteristics of PSF. However, we have to note that only for Gaussian PSF and input, the output given from eq.~(\ref{eq.psfconool}) has the width (variance) equal to the sum of the widths of PSF and input. Normally, on this basis it is possible to link the resolution with the thickness of PSF. In the present analysis we also deal with PSFs very different from a Gaussian function.

Our present analysis is limited to transmission through multilayers surrounded by air and does not include a particular source and detector. An experimental verification of these results is possible in a set-up presented recently by Fabre at.al~\cite{Fabre:PRL-101-073901} for the characterisation of superlenses and schematically shown in fig.~\ref{schematicview}b. A realistic element for use in eg. proximity lithography or for coupling between a plasmonic circuit and an optical fiber requires certain changes to our current assumptions. It would include different materials on both sides of multilayer, and have the input and output planes placed further from the structure. Its design would be preferably based on numerical optimisation of the structure. In this paper instead, we have deliberately chosen to rely on theoretical models represented by the two clearly defined regimes.

\begin{figure}
\begin{center}
\includegraphics[width=7cm]{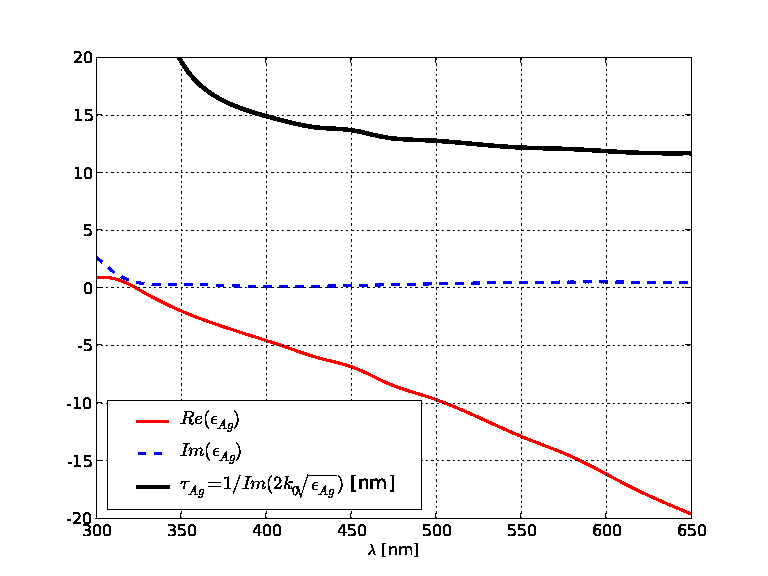}
\end{center}
\caption{Real and imaginary part of the permittivity of silver~\cite{JohnsonChristy}, and the corresponding skin depth.}
\label{eps_ag}
\end{figure}

\subsection{Review of the resonant tunnelling and canalization regimes}
Resonant tunnelling and canalization are the two regimes compared in this paper. Both of them enable a metal-dielectric multilayer to perform the role of a near-field subwavelength imaging device operating at visible wavelengths.

Resonant tunnelling has been proposed and demonstrated already in 1998 by Scalora, Bloemer et.~al~\cite{Scalora1998ap,Bloemer1998apl}, who have shown that in the pass-band of the metal-dielectric stacks around the visible wavelengths, light penetration into metal is not limited to the skin depth and can reach distances of the wavelength order. This phenomenon called \textit{transparency of metals} has been recently recalled~\cite{Scalora2007opex} in the context of near-field subwavelength imaging. Some other interesting features of this regime have been also indicated~\cite{Scalora2007opex}, such as the possibility to tailor the transmission band in a wide range, or the possibility of enlarging the structure with the surprising effect of increasing the transmission. It has also been shown that the transmission depends strongly on the first and last layers which couple and decouple light to and from the periodic structure~\cite{Scalora2007opex}. 

\begin{figure}
\begin{center}
a.\includegraphics[width=7cm]{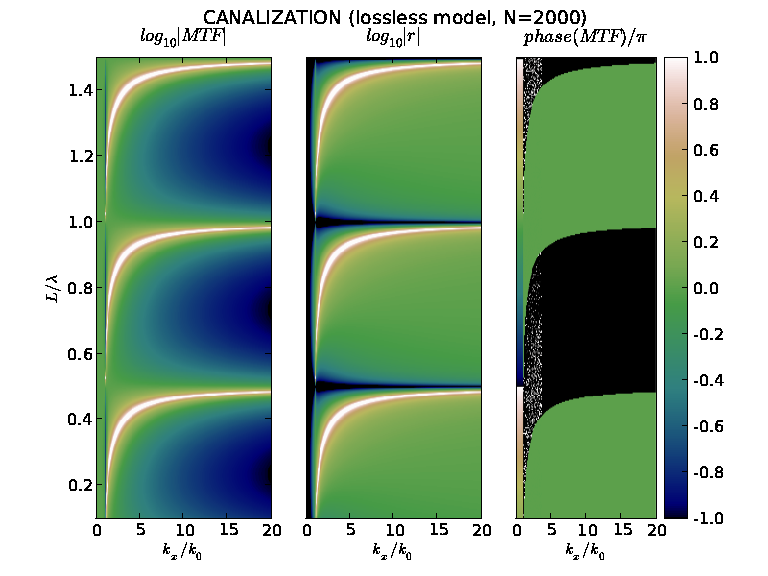}\\
b.\includegraphics[width=7cm]{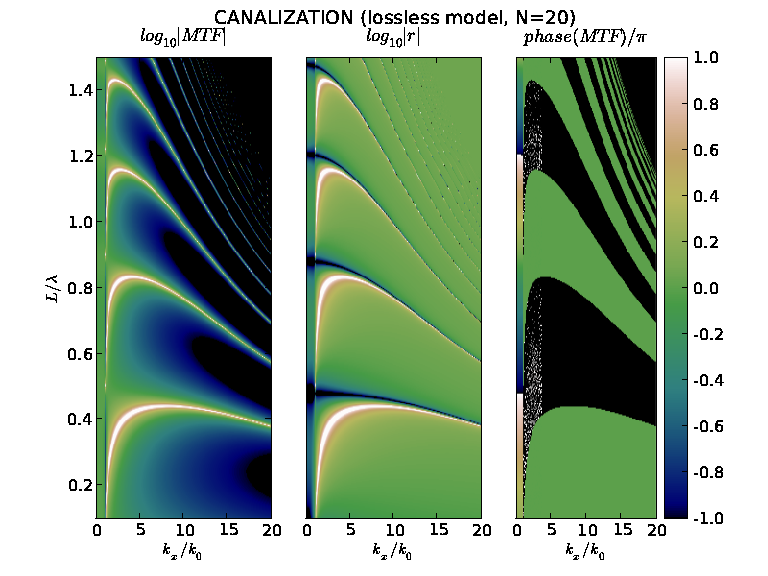}\\
c.\includegraphics[width=7cm]{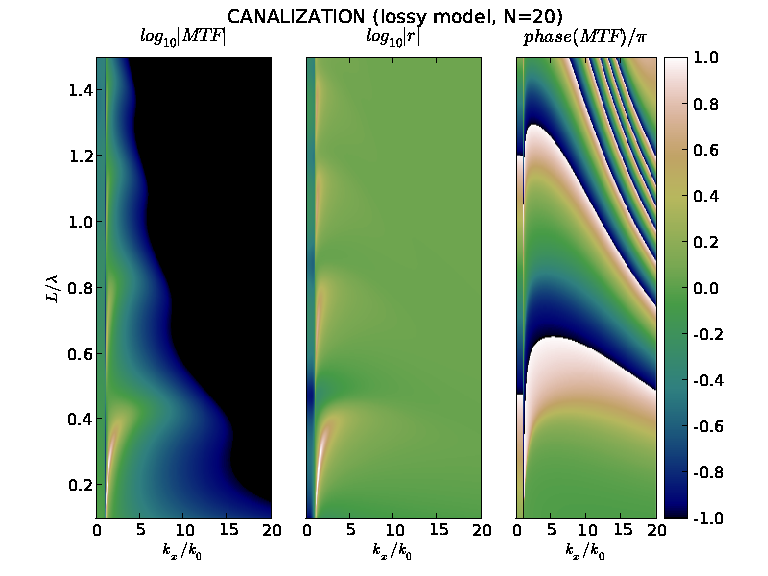}
\end{center}
\caption{MTF and the reflection coefficient for the multilayer with $\epsilon_{x}=1-Re(\epsilon_{Ag})$, $Re(\epsilon_{Ag})/\epsilon_x=-d_{Ag}/d_x$, and $\rho=0$. Resonant condition for canalization is achieved for $2L/\lambda \in Z$. \textbf{a}.~lossless model with a large number of periods $N=2000, \epsilon_{Ag}=-12.88$;  \textbf{b}.~lossless model with a realistic number of periods $N=20, \epsilon_{Ag}=-12.88$; \textbf{c}.~lossy model with a realistic number of periods $N=20, \epsilon_{Ag}=-12.88+0.48\imath$.}
\label{can_mtf}
\end{figure}
 
The second regime to be included in our comparison, named \textit{canalization} has been introduced Belov et.~al.~\cite{belov2005csi} and experimentally demonstated by Ikonen et~al.~\cite{ikonen2006eds} for a capacitively loaded wire medium. In 2006 Belov and Hao~\cite{belov2006prb} extended the concept of canalization to metal-dielectric layered structures. We will only refer to such type of structures in the present paper.
The canalization regime~\cite{belov2006prb} enables subwavelength imaging at visible frequencies through metal-dielectric multilayers. The multilayer consists of layers sufficiently narrow to be properly characterised with the homogenisation predicted by the effective medium theory (EMT). In the canalization regime, the multilayer needs to have the effective permittivity tensor of the anisotropic form with $\epsilon_\bot=\infty$, and $\epsilon_\Vert=1$, and at the same time to fullfill the Fabry-Perot condition for minimal reflections $2L/\lambda\in Z$. This leads to simple geometrical conditions of $\epsilon_{Ag}/\epsilon_x=-d_{Ag}/d_x$ and $\epsilon_x=1-\epsilon_{Ag}$. To fulfil the last condition, one has to neglect the losses in silver, and assume that $\epsilon_{Ag}$, and $\epsilon_x$ are real but have a different sign. In fig.~\ref{eps_ag} we show the dispersion of $\epsilon_{Ag}$ for visible and near-UV wavelengths~\cite{JohnsonChristy}. Indeed the magnitude of imaginary part of $\epsilon_{Ag}$ is significantly lower than the magnitude of its real part, while $1-Re(\epsilon_{Ag})$ falls in the range of most known dielectric materials. Therefore, the canalization regime may be realised with use of silver and a variety of dielectric materials.

\begin{figure}
\begin{center}
a.\includegraphics[width=7cm]{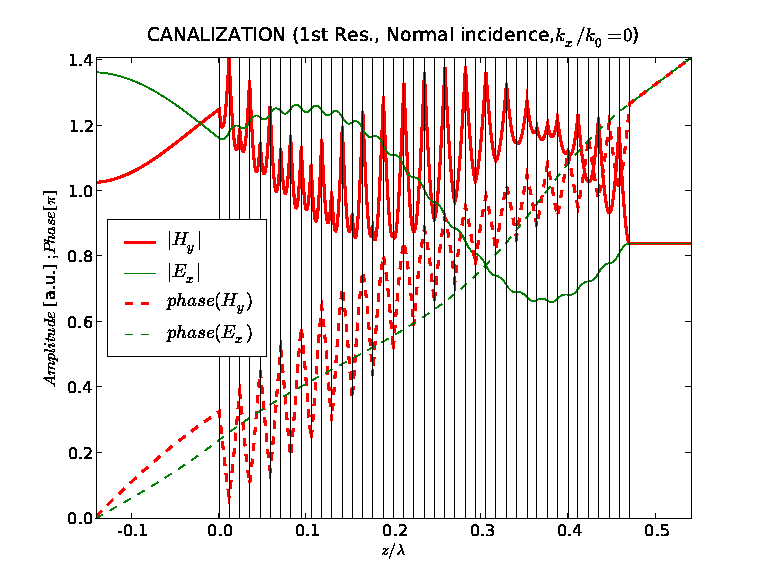}d.\includegraphics[width=7cm]{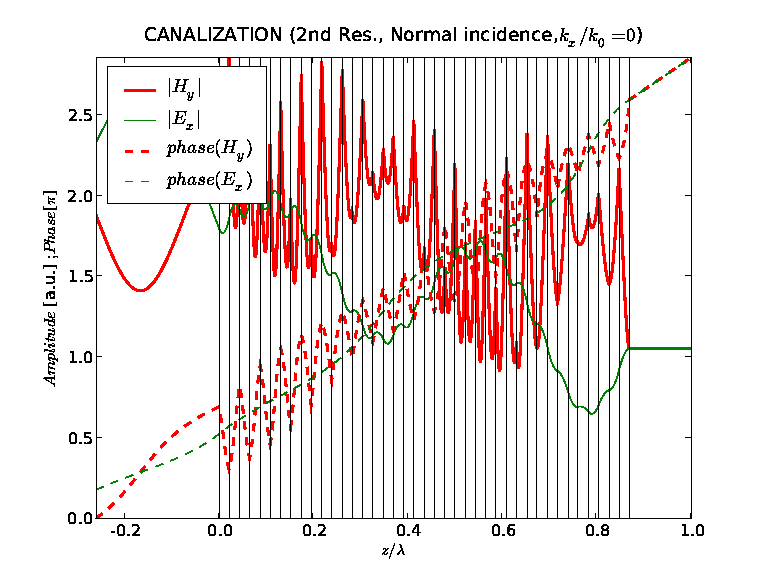}
b.\includegraphics[width=7cm]{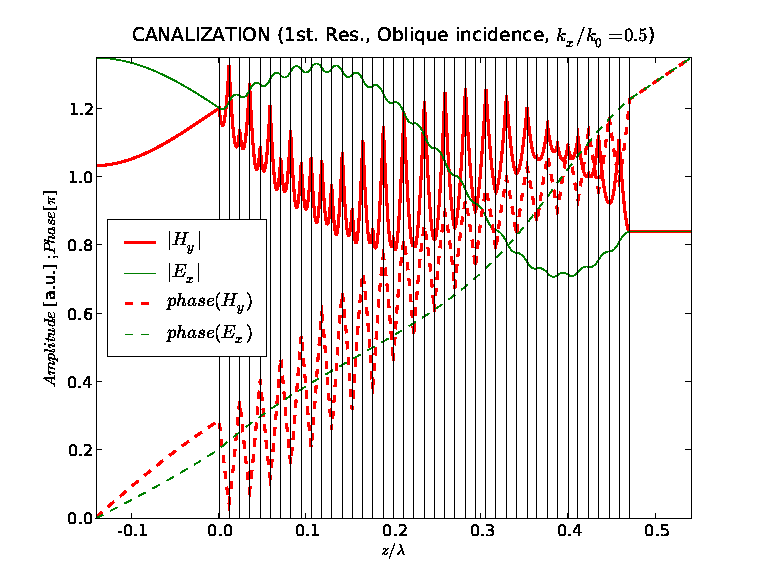}e.\includegraphics[width=7cm]{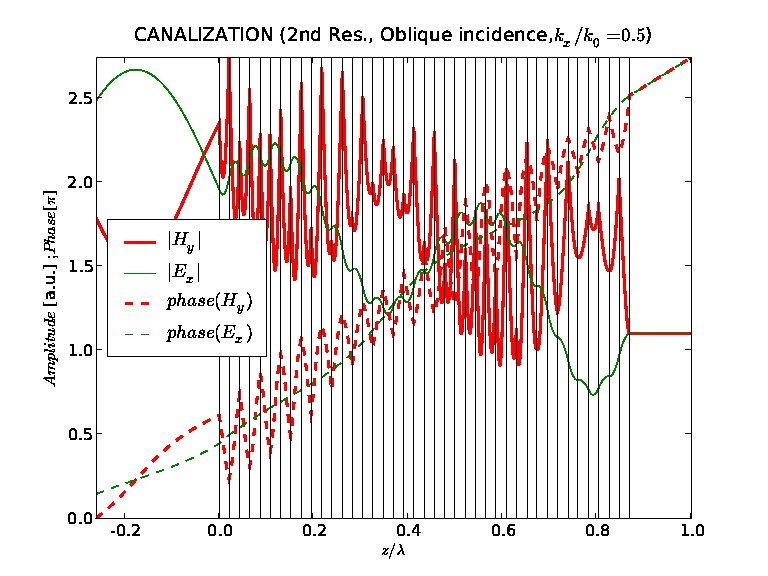}
c.\includegraphics[width=7cm]{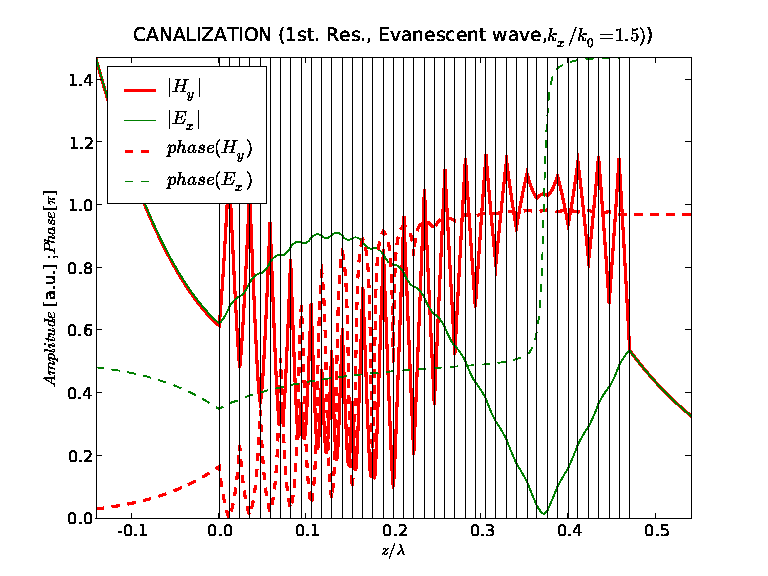}f.\includegraphics[width=7cm]{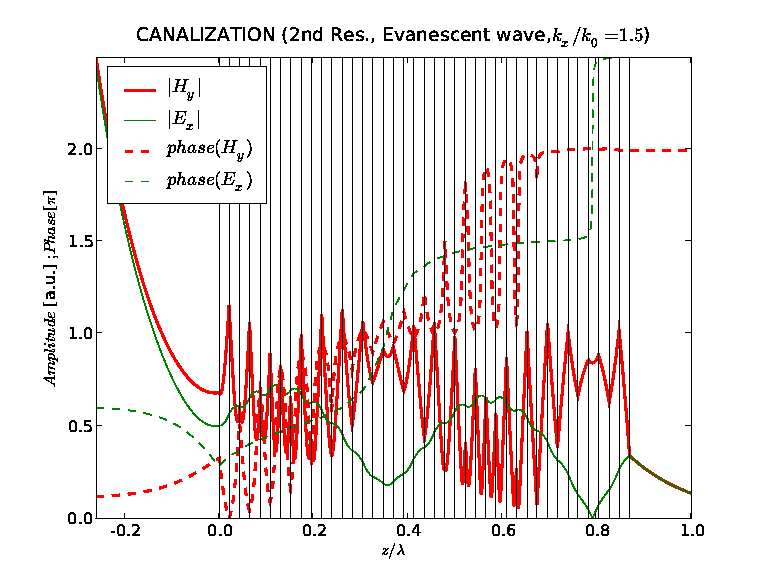}
\end{center}
\caption{Transmission of a plane wave through a metal-dielectric multilayer operating in the canalization regime. \textbf{a,d.}~normal incidence, $k_x=0$;  \textbf{b,e.}~oblique incidence, $k_x=0.5 k_0$; \textbf{c,d.}~evanescent wave, $k_x=1.5 k_0$; Parameters of the multilayer are the same as in fig.~\ref{can_mtf}c. with $L/\lambda=0.47$ (abc) or $L/\lambda=0.87$ (def).}
\label{schem_can}
\end{figure}

\section{Canalization}
Canalization regime can only be achieved when the silver and dielectric layers have real permittivities, and $\epsilon_{Ag}/\epsilon_x=-d_{Ag}/d_x$. Since permittivity of metals is complex, this relation can only be approximately fulfilled. Moreover, the validity of EMT requires that the layers were small with respect to wavelength $\Lambda<<\lambda$.
In this section we investigate the practical importance of both assumptions. For the purpose of this analysis, let us assume that the wavelength is in the middle of the visible range $\lambda=550nm$, when the permittivity of silver is equal to $\epsilon_{Ag}=-12.88+0.48\imath$. The dielectric $x$ needs to have the permittivity of $\epsilon_{x}=13.88$, and $d_{Ag}/d_x\approx0.928$.
In fig.~\ref{can_mtf} we characterise the MTF of such a multilayer, as well as the corresponding reflection coefficient as a function of $L/\lambda$, and $k_x/k_0$. Both amplitudes and phases of the MTF are shown. 
The values of the wavevector component parallel to the boundaries $k_x$ range from  $0$ (normal incidence), through $0<k_x<k_0$ (oblique incidence from air), up to larger values $k_x>k_0$ (evanescent waves) important for sub-wavelength imaging. In fig.~\ref{can_mtf} we compare the MTFs calculated for an unrealistic number of periods $N=2000$  and no losses in metal (fig.~\ref{can_mtf}a), with a more realistic case $N=20$ (fig.~\ref{can_mtf}b), when the EMT applies only approximately. Finally, in fig.~\ref{can_mtf}c, we included losses into the calculation. Canalization requires that $2L/\lambda\in Z$. This is the condition to minimise reflections from the Fabry-Perrot slab with effective anisotropic properties independently of the angle of incidence.
Several conclusions can be drawn from the MTFs in fig.~\ref{can_mtf}. The finite width of the layers primarily affects the high FP resonances, shifts their location towards smaller values of $L/\lambda$ and introduces a certain dependence of the resonant condition on $k_x/k_0$. Further, the transmission bandwidth is reduced by losses (fig.~\ref{can_mtf}c). Moreover, losses introduce a rapid phase variation in the vicinity of $k_x=k_0$, which must result in deteriorated imaging. In practise, only the first two resonances allow for sub-wavelength imaging, therefore limiting the thickness of the slab $L$ to subwavelength distances.

Figs.~\ref{can_mtf}a-c indicated significant differences between the MTFs representing the canalization in its ideally formulated concept, and its implementation with realistic assumptions. Nevertheless, the distribution of field transmitted through the structure (see fig.~\ref{schem_can}a-f) proves that also the realistic imaging device operates according to expectations. In fact, a plane wave under normal incidence (fig.~\ref{schem_can}ad), oblique incidence (fig.~\ref{schem_can}be), and an evanescent wave (fig.~\ref{schem_can}cf) are all transmitted with high efficiency. Moreover, the 
distributions of $H_y$ and $E_x$ inside the structure, all represent the same FP resonant condition of $L/\lambda\approx0.5$ or $L/\lambda\approx1$, which should not depend on the angle of incidence. For these suboptimal but realistic conditions, both for the propagating and evanescent waves, the multilayer is therefore characterised with a similar Bloch vector, and operates as a FP plate at resonant conditions independently on $k_x/k_0$.
On average, the phase of both $H_y$ and $E_x$ increases in the direction of propagation for the propagating waves with the same slope as in air, indicating the same effective phase velocity. However, locally the phase of $H_y$ propagates backwards in the silver layers.

\begin{figure}
\begin{center}
a.\includegraphics[width=7cm]{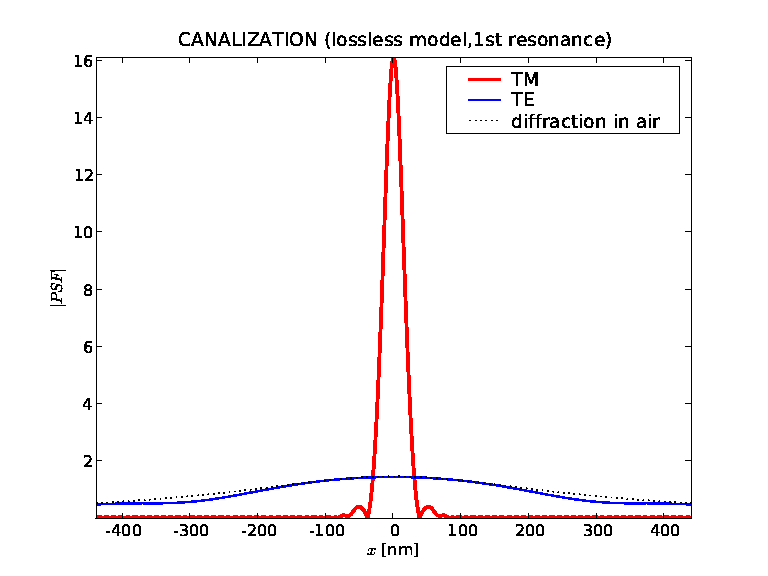}  
b.\includegraphics[width=7cm]{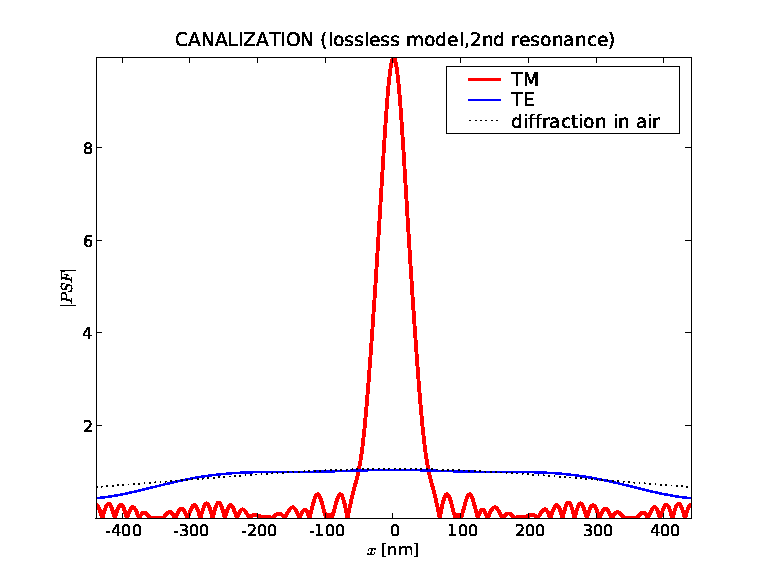}
\end{center}
\caption{PSF of the lossless multilayer with same parameters as in fig.~\ref{can_mtf}b. operating in the canalization regime \textbf{a}.~at the $L/\lambda=0.48$ resonance; \textbf{b}.~at the $L/\lambda=0.88$ resonance. For comparison, the PSF for the TE-polarisation and for the diffraction in air at the same distance are also shown.}
\label{can_psf_lossless}
\end{figure}

\begin{figure}
\begin{center}
a.\includegraphics[width=7cm]{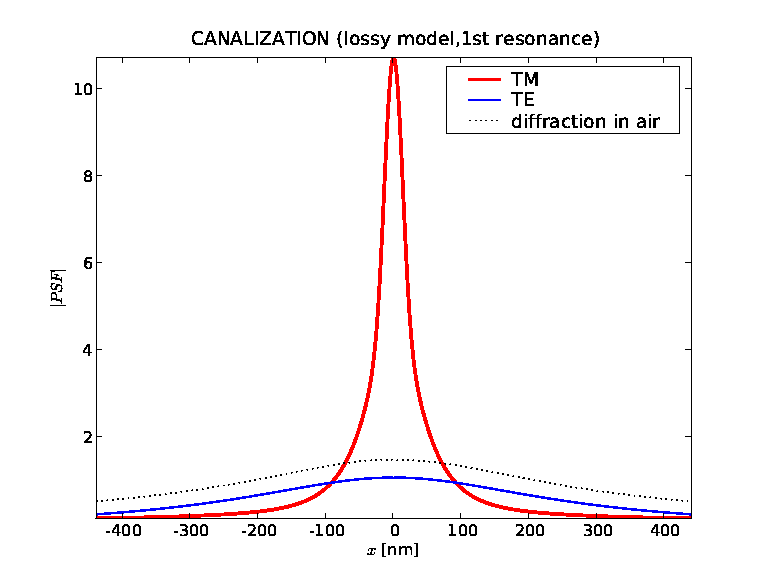}
b.\includegraphics[width=7cm]{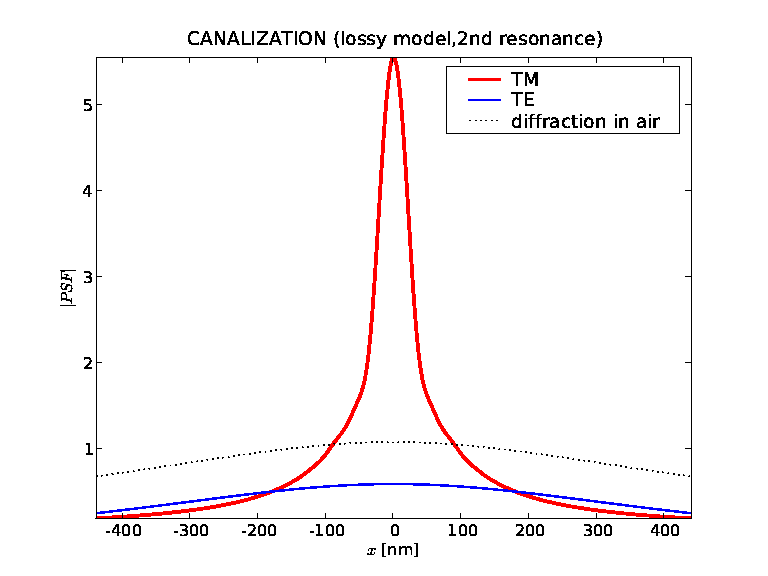}
\end{center}
\caption{PSF of the lossy multilayer with the same parameters as in fig.~\ref{can_mtf}c operating in the canalization regime  \textbf{a}.~at the $L/\lambda=0.47$ resonance; \textbf{b}.~at the $L/\lambda=0.87$ resonance. For comparison, the PSF for the TE-polarisation and for the diffraction in air at the same distance are also shown.}
\label{can_psf_lossy}
\end{figure} 

Resolution of the imaging system normally can be deduced from the width of the PSF. In fig.~\ref{can_psf_lossless} we present the PSF corresponding to the first and second FP resonances, when the losses are neglected. In order to determine the actual location of the resonances we minimised the ratio between reflection and transmission for normal incidence.  PSF is calculated from the MTF in fig.~\ref{can_mtf}b., for $\lambda=550nm$. In Table~\ref{tab.resolution} we compare the FWHM and standard deviation of the PSFs presented in this paper. The thickness of the multilayer is equal to $L=0.48\lambda=264nm$, or $L=0.88\lambda=484nm$, for the two resonances respectively. The central maximum of the $|PSF|^2$ has the FWHM equal to $23nm$ or $29nm$ (see Table~\ref{tab.resolution}), allowing for imaging with sub-wavelength resolution. The PSFs for free-space propagation along the same distances are much broader (see figs.~\ref{can_psf_lossless}ab). The PSF for the first resonance is approximately Gaussian in shape. PSF for the second resonance (fig.~\ref{can_mtf}b) has approximately twice lower amplitude and is distorted by the presence of a background structure resulting from a SPP mode that appeared for large $k_x/k_0$ in the MTF. In the presence of losses, this mode has been suppressed (see fig.~\ref{can_mtf}c) resulting in the regularisation of the corresponding PSF. Figure~\ref{can_psf_lossy}ab show the PSF for the first two FP resonances, with losses included into calculation. The apparent regularisation of the PSF by the presence of losses appeared at the price of reduced amplitude and a slower decay far from the center (this is also visible from the large values of standard deviation in comparison to FWHM in Tab.\ref{tab.resolution}). However, the resolution measured with FWHM is only little affected. 

We consider the approximately Gaussian shape of the PSF as a major advantage of the canalization regime.
This highly desired feature reflects the regular shape of the MTF. In particular, imaging in the canalization regime does not involve plasmons, and for $L/\lambda=0.5$, the MTF does not experience significant enhancements for evanescent waves. Moreover, its phase behaves smoothly near $k_x/k_0\approx1$. In the lossless case, phase of the MTF is approximately constant in the neighbourhood of that point. Losses affect this situation only in the very close vicinity of $k_x/k_0\approx1$, and a local phase modulation around that point has no major influence on the shape of PSF.
However, due to the way in which losses affect the MTF, it will be difficult to utilise higher order FP resonances, and therefore construct multilayers thicker than the wavelength.

\section{Resonant tunnelling - broadband transparency}

\begin{figure}
\begin{center}
a.\includegraphics[width=7cm]{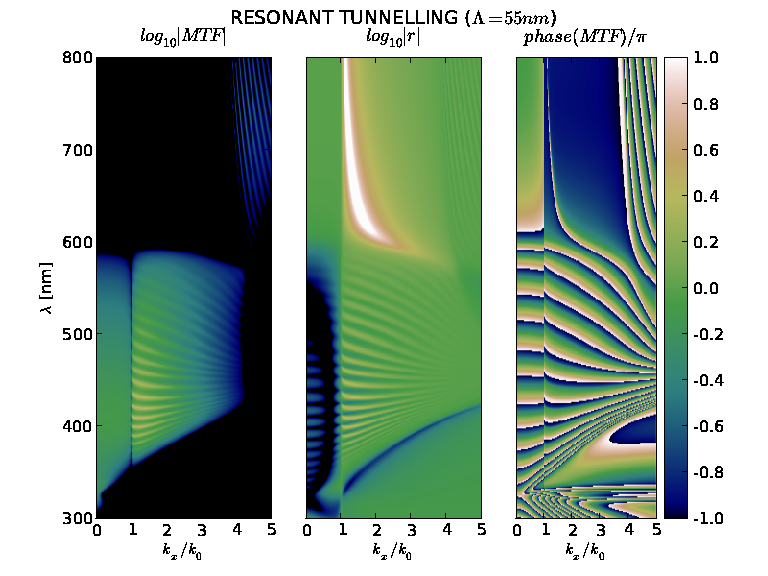}
b.\includegraphics[width=7cm]{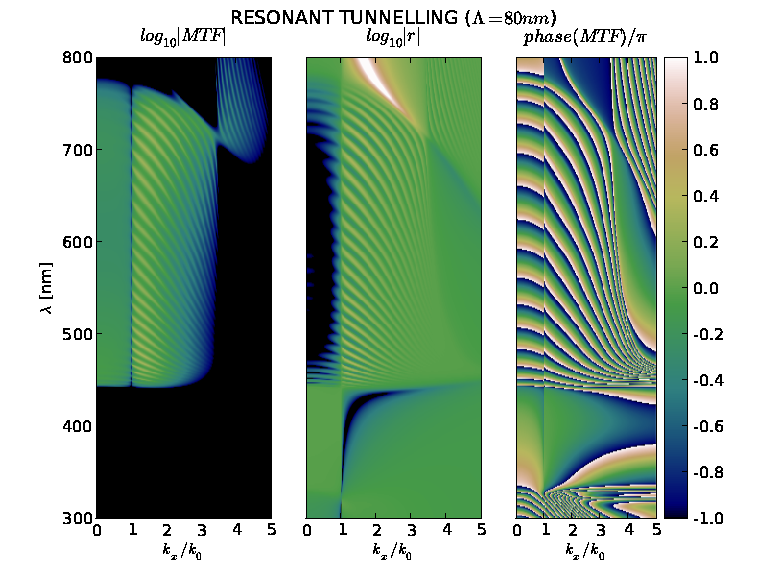}
\end{center}
\caption{MTF of the multilayer operating in the resonant tunnelling regime described by the following parameters: $N=30, \rho=0.5,  d_{Ag}=20nm, \epsilon_x=8$ and  \textbf{a}.~$\Lambda=55nm$; \textbf{b}.~$\Lambda=80nm$.}
\label{tun_mtf}
\end{figure} 

\begin{figure}
\begin{center}
a.\includegraphics[width=7cm]{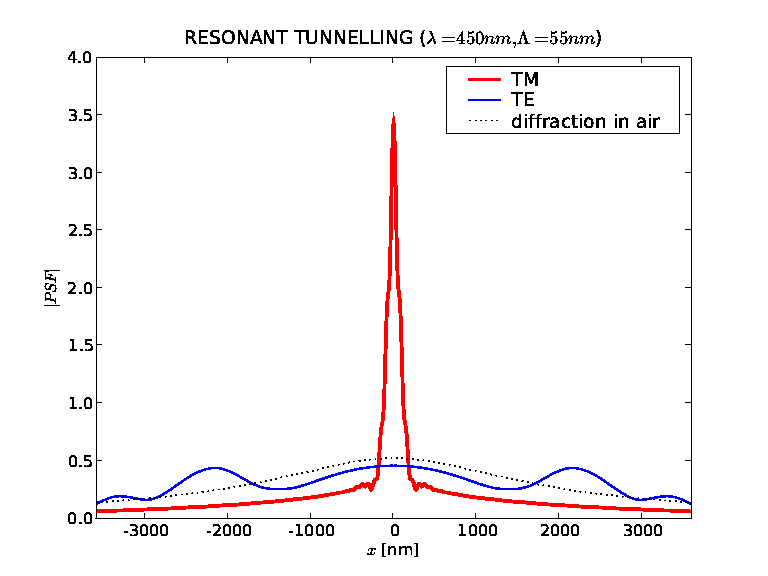}b.\includegraphics[width=7cm]{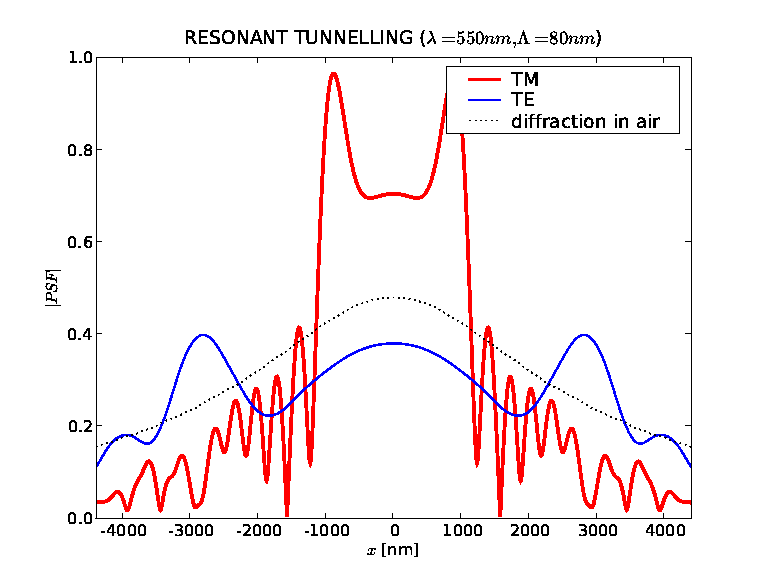}
\end{center}
\caption{PSF of the multilayer operating in the resonant tunnelling regime \textbf{a.} for the same multilayer as in fig.~\ref{tun_mtf}a, at $\lambda=450nm$; \textbf{a.} for the same multilayer as in fig.~\ref{tun_mtf}b, at $\lambda=550nm$; PSF is shown for the TM, and TE polarisations as well as for the diffraction in air at the same distance.}
\label{tun_psf}
\end{figure} 

\begin{figure}
\begin{center}
a.\includegraphics[width=7cm]{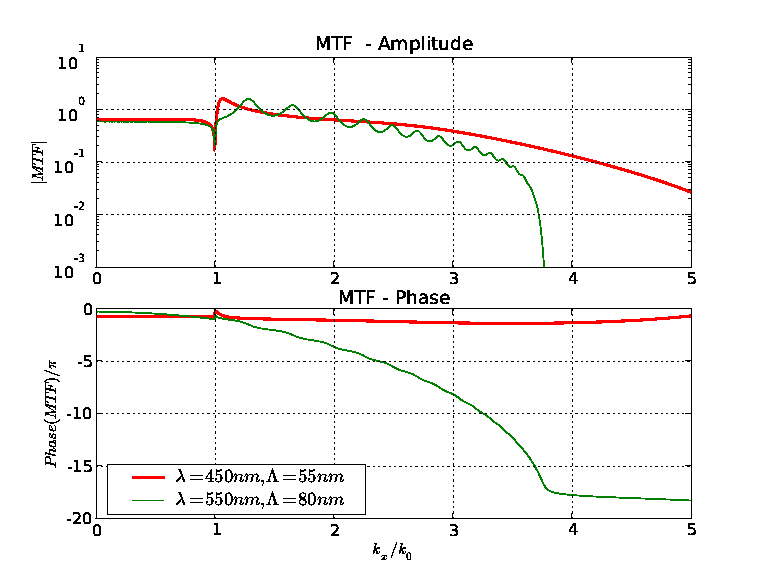}
b.\includegraphics[width=7cm]{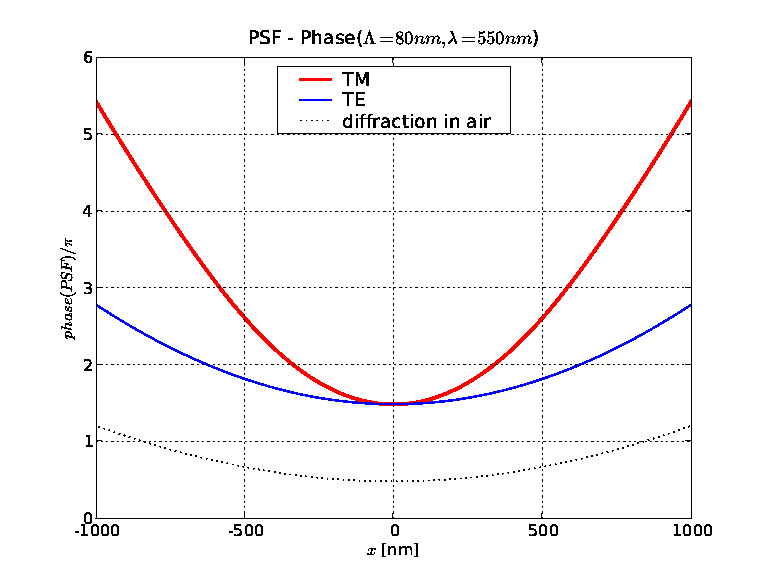}
\end{center}
\caption{\textbf{a.}Amplitude and phase of the MTF of the multilayers operating in the resonant tunnelling regime from fig.~\ref{tun_mtf}a (cross-section at $\lambda=450nm$) and fig.~\ref{tun_mtf}b (cross-section at $\lambda=550nm$);  \textbf{b.} Phase of the central part of PSF from fig.~\ref{tun_psf}b.}
\label{tun_psf_80etc}
\end{figure}

\begin{figure}
\begin{center}
a.\includegraphics[width=7cm]{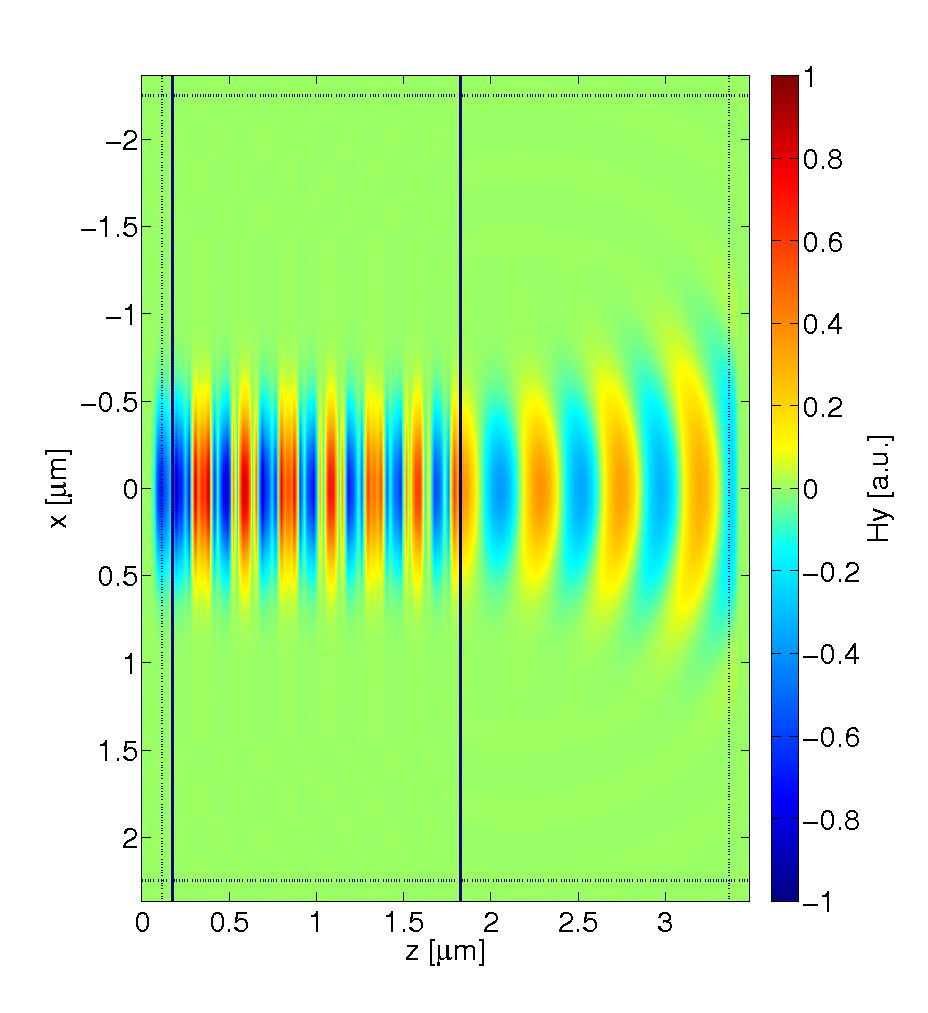}
b.\includegraphics[width=7cm]{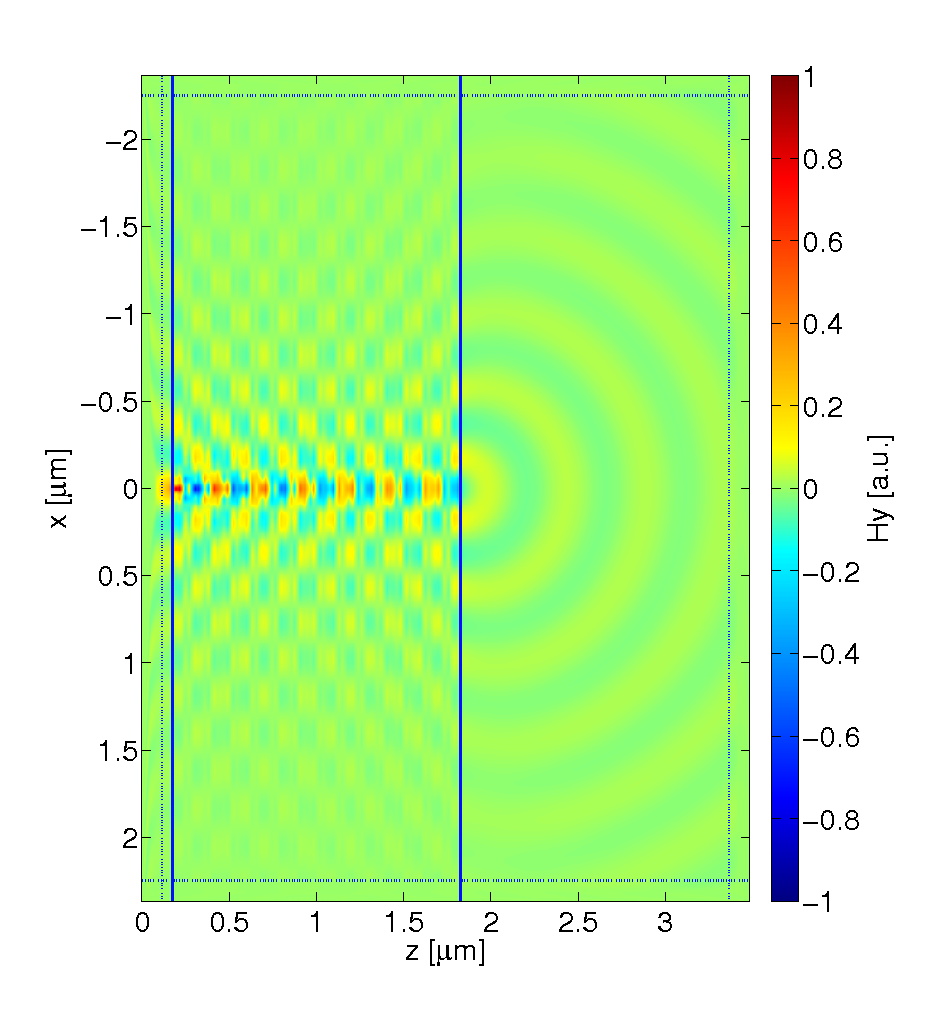}
c.\includegraphics[width=7cm]{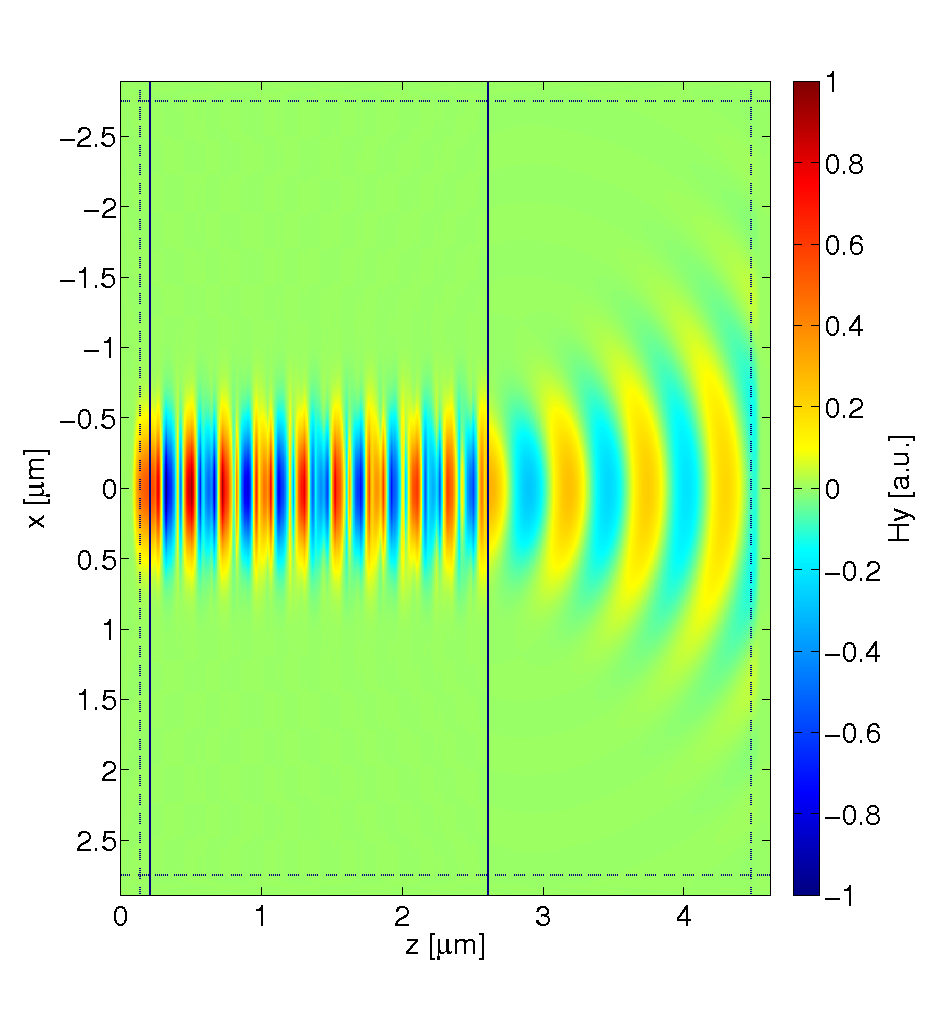}
d.\includegraphics[width=7cm]{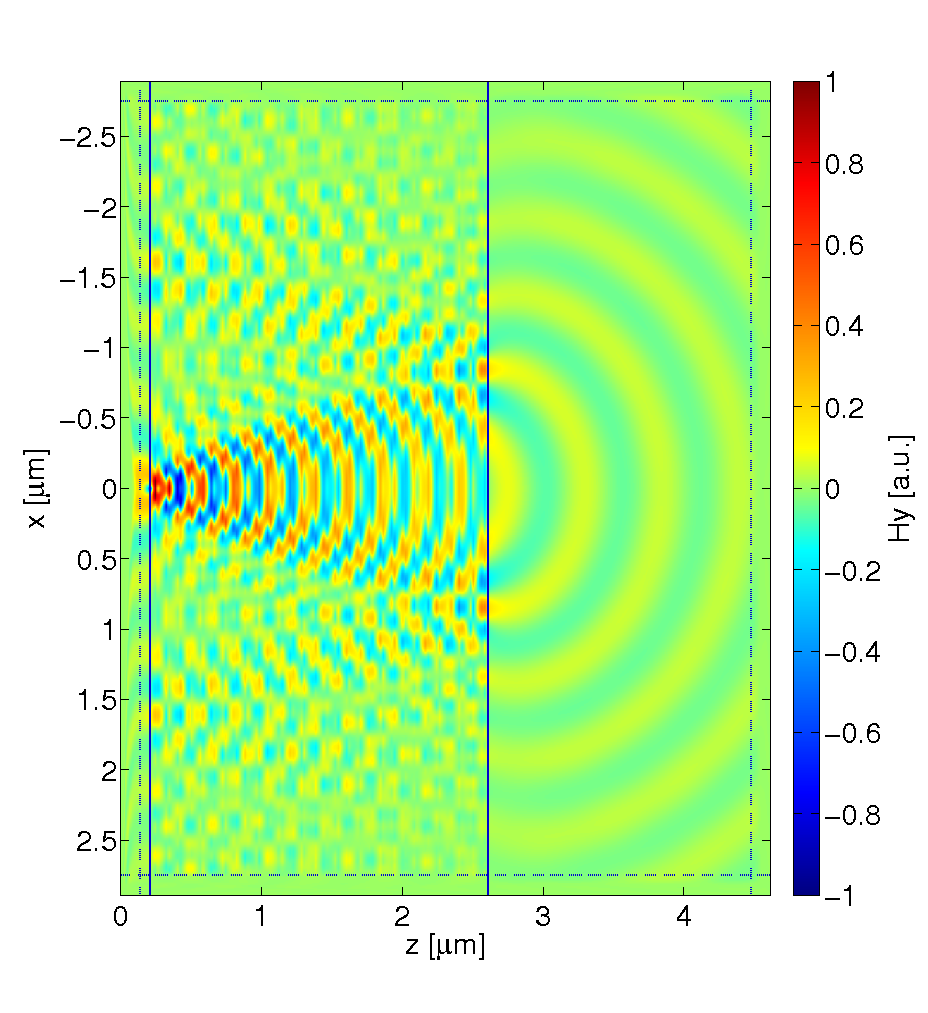}
\end{center}
\caption{Resonant tunnelling regime: \textbf{ab}.  $\Lambda=55nm$ $\lambda=450nm$; \textbf{cd}. $\Lambda=80nm$ $\lambda=550nm$. Figures include the instantaneous distribution of $H_y$ for CW illumination in the steady-state, obtained using FDTD when the source is a Gaussian beam (ac), or a point source~(bd) put $5nm$ away from the surface of the multilayer.}
\label{fdtd_tun}
\end{figure}

\begin{figure}
\begin{center}
a.\includegraphics[width=7cm]{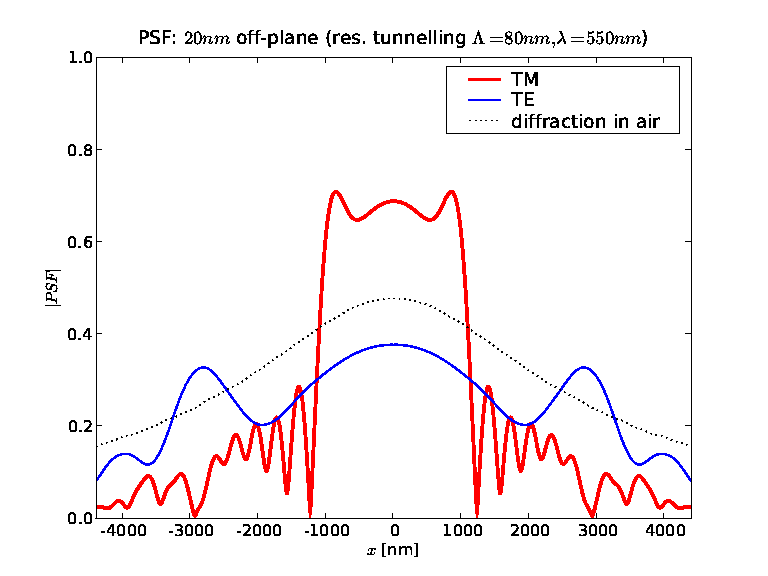}
b.\includegraphics[width=7cm]{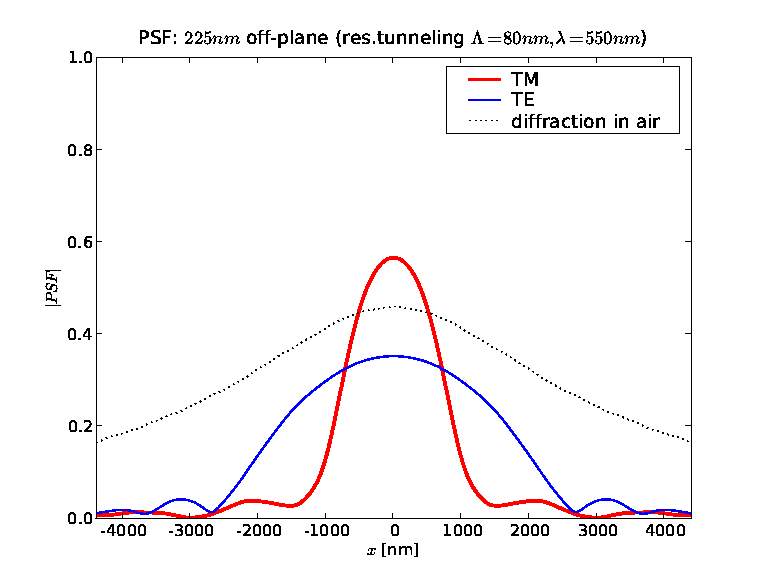}
\end{center}
\caption{\textbf{a}.~PSF of the multilayer operating in the resonant tunnelling regime described by the same parameters as in fig.~\ref{tun_psf}b, however with the position of the incidence plane shifted from the boundary of the structure \textbf{a.}~by $20nm$ \textbf{a.}~by $225nm$.}
\label{tun_psf_detune}
\end{figure} 

\begin{figure}
\begin{center}
a.\includegraphics[width=7cm]{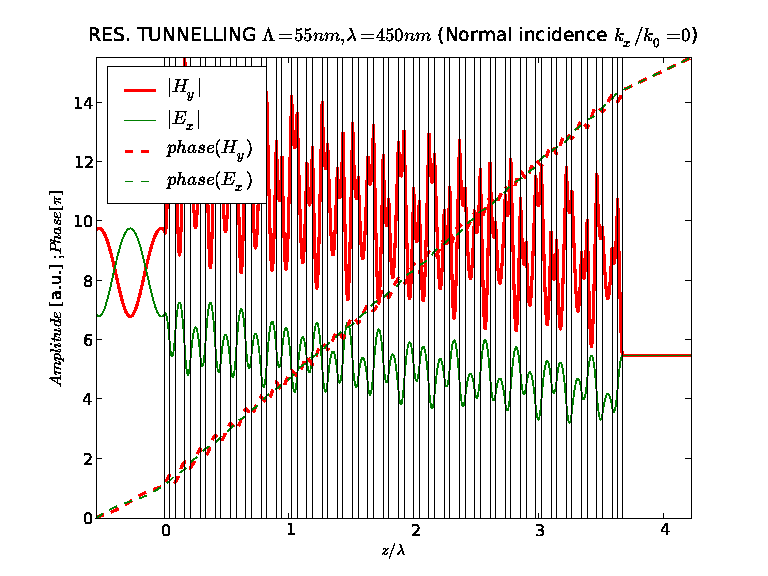}d.\includegraphics[width=7cm]{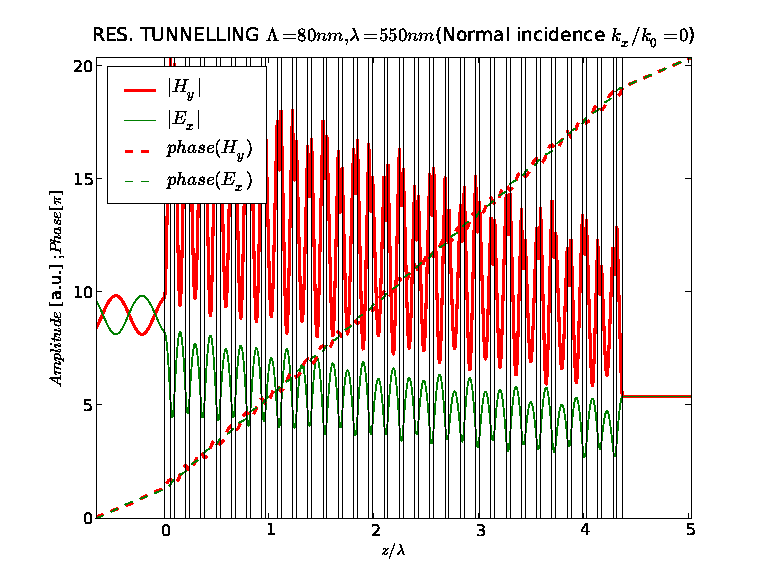}
b.\includegraphics[width=7cm]{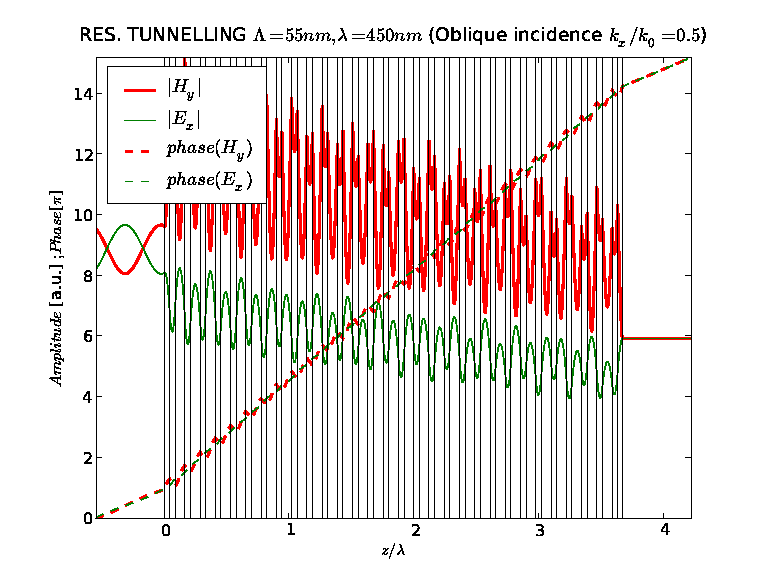}e.\includegraphics[width=7cm]{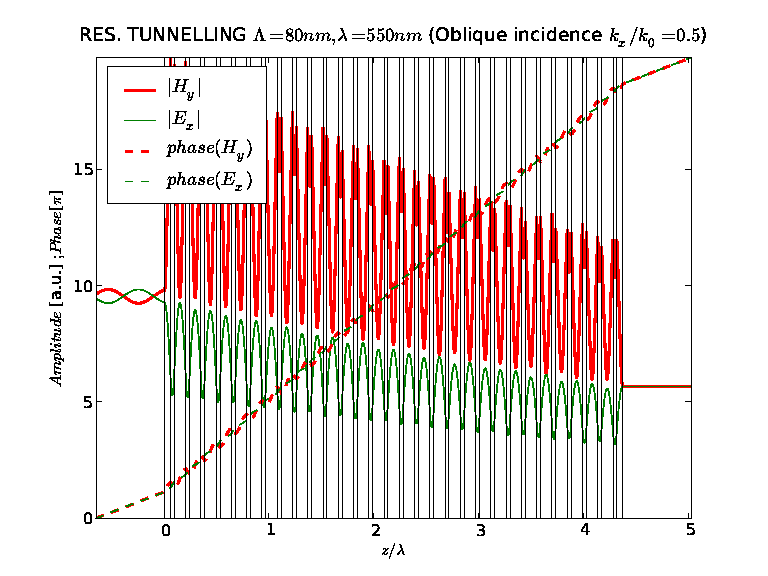}
c.\includegraphics[width=7cm]{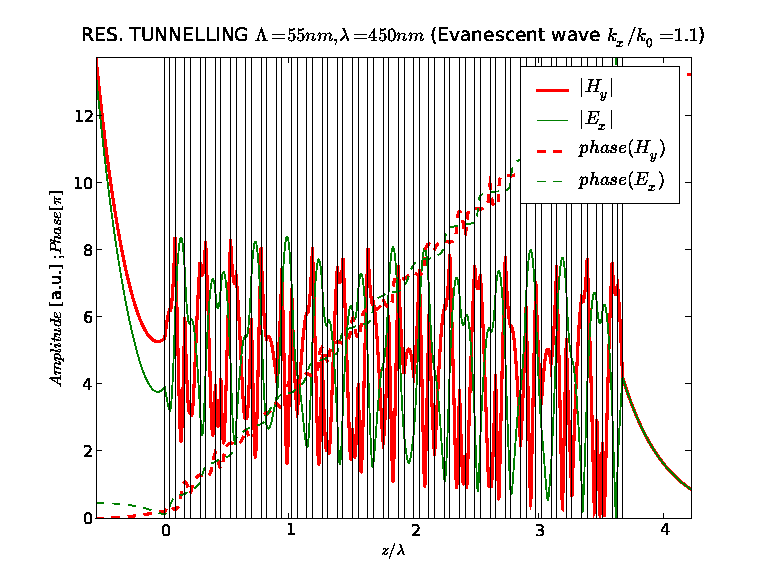}f.\includegraphics[width=7cm]{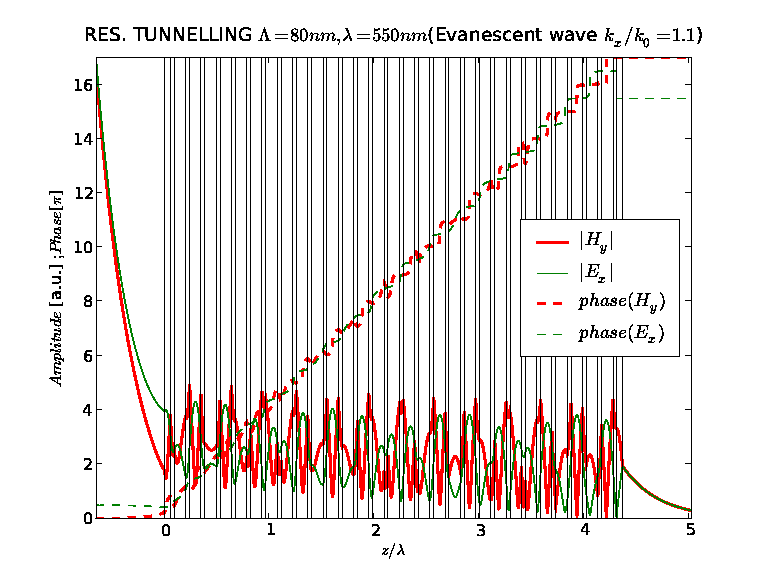}
\end{center}
\caption{Transmission of a plane wave through the multilayer operating in the resonant tunnelling regime. \textbf{a,d.}~normal incidence, $k_x/k_0=0.$;  \textbf{b,e.}~oblique incidence, $k_x/k_0=0.5$; \textbf{c,f.}~evanescent wave, $k_x/k_0=1.1$; Parameters of the multilayer are the same as in fig.~\ref{tun_mtf}a. for $\lambda=450nm$ (abc) or in fig.~\ref{tun_mtf}b. for $\lambda=550nm$ (def). }
\label{schem_tun}
\end{figure} 

We begin by stating the known differences between canalization and resonant tunnelling. Resonant tunnelling~\cite{Scalora1998ap,Scalora2007opex}, as opposed to canalization, allows for broadband transmission of light. The transmission window may be tuned in a large range and corresponds to the pass-band of the photonic crystal. Transmission has no straightforward dependence on homogenisation, and the layers may be thicker than in the canalization regime. However, the first and last layers perform a role similar to anti-reflection coatings and have a major influence on the transmission. They couple the incident wave to a Bloch wave propagating through the periodic structure, and decouple it in the end. No special relation between the permittivity of metal and dielectric components is required, leaving much space for numerical optimisation.

We focus on the imaging properties of multilayers operating under resonant tunnelling. Let us consider a multilayer consisting of silver
and a dielectric with an arbitrarily selected permittivity of $\epsilon_x=8$.
In fig.~\ref{tun_mtf} we present the wavelength dependence of the MTF and reflection coefficient for two multilayers. Both include $N=30$ periods, with $Ag$ layers thickness close to the skin depth $d_{Ag}=20nm$, and either $d_x=35nm$, or $d_x=60nm$.
The losses and dispersion of silver are taken into account.
Following the suggestion from~\cite{Scalora2007opex}, we have chosen the external layers to have the thickness such that $\rho=0.5$. This value optimised the transmission in our simulations, as well.
We emphasise that the general characteristics of the MTFs obtained with a wide range of different parameters are very similar to those analysed in fig.~\ref{tun_mtf}. These MTFs are typical for the resonant tunnelling. 

The results from fig.~\ref{tun_mtf}ab indicate the existence of a broad spectral region of transparency at $350nm-550nm$ or $450nm-760nm$, respectively. Tuning to the desired range is based on the simple choice of thicknesses $d_x$.
Transparency windows extend to high evanescent waves $k_x/k_0\sim3$, however with certain restrictions. Transmission of evanescent waves is accompanied by high reflections. The corresponding MTF is not a uniform function of $k_x/k_0$ and experiences several maxima and a phase modulation. Finally the phase is not continuous at $k_x/k_0=1$. These features may result in a distorted shape of PSFs as compared to the previously discussed canalization regime.

In fig.~\ref{tun_psf}a and fig.~\ref{tun_psf}b we present two selected PSFs for the resonant tunnelling regime. PSF form fig.~\ref{tun_psf}a
corresponds to the MTF from fig.~\ref{tun_mtf}a calculated for the wavelength of $\lambda=450nm$,
while PSF form fig.~\ref{tun_psf}b corresponds to the MTF from fig.~\ref{tun_mtf}b calculated for the wavelength of $\lambda=550nm$. 
Both PSFs are significantly narrower compared to the corresponding PSF for propagation in air, or to the PSFs for the TE polarisation. While, in its centre the PSF in fig.~\ref{tun_psf}a resembles a narrow Gaussian function, it also contains a slowly decaying backround. The PSF in fig.~\ref{tun_psf}b has an interesting structure with two maxima and a periodic modulation. The astonishing difference between PSFs from fig.~\ref{tun_psf}a and  fig.~\ref{tun_psf}b results from the very different phase slopes of the corresponding MTFs, which we compare in fig.~\ref{tun_psf_80etc}a. The broadening of PSF effects from the large slope of phase in the evanescent part of the spatial spectrum.

The two maxima of PSF from fig.~\ref{tun_psf}b are spatially separated by around $3\lambda$. Such a shape of PSF could indicate the possibility of using the considered structure as a beam-splitter. However, this PSF is not a simple superposition of two Gaussian-like functions but also is subject to a rapid phase modulation (see fig.\ref{tun_psf_80etc}b). In effect it transforms the incident wavefront in a way which is different than simple splitting and broadening. To illustrate this, we show in fig~\ref{fdtd_tun} how the corresponding multilayer transmitts a narrow Gaussian beam comparing it to the transmission of a beam incident from a point source. This simulation is based on the finite difference time domain method~\cite{Taflove:FDTD} (FDTD). FDTD Simulations were performed using a freely available MEEP software package with subpixel smoothing for increased accuracy\cite{Farjadpour2006ol}. The TMM is a quasi-analytical method, while accurate FDTD simulations of metallic layered structures are difficult~\cite{zhao2007prb} and may not reproduce the fine details of the field. Nevertheless our FDTD simulations come in good agreement with TMM. 

The conditions for the resonant tunnelling regime correspond to those in fig.~\ref{tun_psf}ab. with a CW source put very close ($\approx 5nm$) to the surface of the multilayer.
In contrast to the point source, the Gaussian beam is neither split nor broadened (figs.~\ref{fdtd_tun}cd). 

This difference deserves a further explanation. 
The decomposition of PSF from fig.~\ref{tun_psf}b into spatial harmonics differs significantly from that of a Gaussian function with a similar width. The Gaussian beam is composed mainly from the propagating spatial harmonics, while the broad shape of PSF (fig.~\ref{tun_psf}b) results from the rapid phase modulation in the evanescent part of the spatial spectrum (fig.~\ref{tun_psf_80etc}a.).
For similar reasons, the PSF from fig.~\ref{tun_psf}b is also highly sensitive to the experimental conditions. For instance, an additional propagation in air either before or after the multilayer changes the shape of the PSF significantly by filtering out the evanescent part of the MTF. In fig.~\ref{tun_psf_detune}
we show how the PSF is modified with distance. Already after $20nm$ the side-maxima are attenuated to the size of the central maximum (fig\ref{tun_psf_detune}a). After the distance of $225nm$ the PSF becomes approximately Gaussian. Its width is reduced to the central maximum providing a qualitative agreement with FDTD simulation for the Gaussian incidence.

For completeness, in fig.~\ref{schem_tun} we present the internal distribution of electric and magnetic field for a plane wave transmitted through the multilayer. The multilayer operating in the resonant tunnelling regime is the same as in figs.~\ref{tun_mtf}b, and ~\ref{tun_psf}b. Fig.~\ref{schem_tun}a,b, and c, show respectively the transmission of a normally incident, inclined or evanescent wave, through the whole structure. In contrast to fig.~\ref{schem_can} for the canalization regime, there is no indication of a FP resonance this time. Indeed both transmission mechanisms are different, and as we show provide different PSF characteristics.

\begin{table}
\caption{Resolution in the canalization and resonant tunnelling regimes. Emphasised values indicate the PSF widths below $\lambda/2$ and structure thicknesses over $\lambda$.}
\label{tab.resolution}
\begin{tabular}
{ | p{7cm} | l | l | l |l| l|}
\hline 
Description & Fig. & $\lambda$ [nm] & $\frac{FWHM_{|PSF|^2}}{\lambda}$ & $\frac{\sigma_{|PSF|^2}}{\lambda}$ & $\frac{L}{\lambda}$ \\
\hline
Canalization, lossless, 1st resonance & \ref{can_psf_lossless}a & $550$ & $\textbf{0.042}$ & $\textbf{0.019}$ & $0.5$ \\
Canalization, lossless, 2nd resonance & \ref{can_psf_lossless}b & $550$ & $\textbf{0.066}$ & $\textbf{0.049}$ & $\textbf{1.0}$ \\
Canalization, lossy, 1st resonance & \ref{can_psf_lossy}a & $550$ & $\textbf{0.052}$ & $\textbf{0.060}$ & $0.47$ \\
Canalization, lossy, 2nd resonance & \ref{can_psf_lossy}b & $550$ & $\textbf{0.074}$ & $\textbf{0.11}$ & $0.87$ \\
\hline
Res. tunnelling $\Lambda=55nm$& \ref{tun_psf}a & $450$ & $\textbf{0.17}$ & $0.99$ & $\textbf{3.67}$ \\
Res. tunnelling $\Lambda=80nm$& \ref{tun_psf}b & $550$ & $3.89$ & $1.76$ & $\textbf{4.36}$ \\
$20nm$ off-plane & \ref{tun_psf_detune}a & $550$ & $3.85$ & $1.56$ & $\textbf{4.40}$ \\
$225nm$ off-plane& \ref{tun_psf_detune}b & $550$ & $2.18$ & $0.85$ & $\textbf{4.77}$ \\
\hline
\end{tabular}
\end{table}

\section{Conclusion}
We compare subwavelength imaging in the canalization~\cite{belov2006prb} and resonant tunnelling~\cite{Scalora2007opex} regimes.

The canalization regime is based on the homogenisation of the multilayer and enables transmission for resonant frequencies or slab widths. The conditions for canalization can not be accurately achieved in presence of losses. We show that in the lossless case, the phase of MTF is approximately constant in a wide range of $k_x/k_0$. Losses affect this situation mainly in the very close vicinity of $k_x/k_0\approx 1$, and a local phase modulation around that point has no major influence on the shape of PSF. Losses primarily decrease the amplitude of the PSF and suppress the possibility of using higher FP resonances. Losses also remove the background noise in the PSF, and slightly affect the resolution. In effect, the PSF is approximately Gaussian in shape, which is a major advantage of the canalization regime.

Resonant tunnelling enables broadband transmission through multilayers with thickness ranging up to several wavelengths. We demonstrate that the resulting PSFs may be distorted by sidelobes, and slow attenuation far from the central position. These effects are not entirely suppressed by losses. Such behaviour results from the shape of corresponding MTF, which is not a uniform function of $k_x/k_0$, experiences multiple maxima and a phase modulation, and finally the phase of MTF is not continuous at $k_x/k_0=1$. Nevertheless, subwavelength imaging is still possible. We demonstrate a situation, when the PSF becomes narrower with growing distance from the point source. This effect may be understood from the rapid phase modulation of both MTF and  PSF. Further from the multilayer, the evanescent part of MTF with rapid phase modulation is filtered out, and the PSF becomes narrower and gains a regular Gaussian shape.

\section*{Acknowledgement}
 We acknowledge support from the Polish MNiI research project
{0694/H03/ 2007/32} and the frameworks of {COST~MP0702} and {MP0803} actions.

\section*{References}
\bibliographystyle{unsrt}
\bibliography{kotynski_stefaniuk}

\begin{thebibliography}{10}

\bibitem{pendry2000nrm}
J.~B. Pendry.
\newblock Negative refraction makes a perfect lens.
\newblock {\em Phys. Rev. Lett.}, 85(18):3966--3969, 2000.

\bibitem{anantharamakrishna2002aln}
S.~A. Ramakrishna, J.~B. Pendry, D.~Schurig, and D.~R. Smith.
\newblock The asymmetric lossy near-perfect lens.
\newblock {\em J. Mod. Opt.}, 49(10):1747--1762, 2002.

\bibitem{melville2005sri}
D.~O. Melville and R.~J. Blaikie.
\newblock Super-resolution imaging through a planar silver layer.
\newblock {\em Opt. Express}, 13(6):2127--2134, 2005.

\bibitem{fang2005sdl}
N.~Fang, H.~Lee, C.~Sun, and X.~Zhang.
\newblock Sub-diffraction-limited optical imaging with a silver superlens.
\newblock {\em Science}, 308(5721):534--537, 2005.

\bibitem{notomi2000tlp}
M.~Notomi.
\newblock Theory of light propagation in strongly modulated photonic crystals:
  Refractionlike behavior in the vicinity of the photonic band gap.
\newblock {\em Phys. Rev. B}, 62(16):10696--10705, 2000.

\bibitem{Veselago68}
V.~G. Veselago.
\newblock The electrodynamics of substances with simultaneously negative values
  of permittivity and permeability.
\newblock {\em Sov. Phys. Usp.}, 10:509, 1968.

\bibitem{Fabre:PRL-101-073901}
N.~Fabre, L.~Lalouat, B.~Cluzel, X.~Melique, D.~Lippens, F.~de~Fornel, , and
  O.~Vanbesien.
\newblock Optical near-field microscopy of light focusing through a photonic
  crystal flat lens.
\newblock {\em Phys. Rev. Lett.}, 101:073901, 2008.

\bibitem{webb2004mnr}
K.~J. Webb, M.~Yang, D.~W. Ward, and K.~A. Nelson.
\newblock Metrics for negative-refractive-index materials.
\newblock {\em Phys. Rev. E}, 70(3):35602, 2004.

\bibitem{smith2002lsd}
D.~R. Smith, D.~Schurig, M.~Rosenbluth, S.~Schultz, S.~A. Ramakrishna, and
  J.~B. Pendry.
\newblock Limitations on subdiffraction imaging with a negative refractive
  index slab.
\newblock {\em Appl. Phys. Lett.}, 82:1506, 2003.

\bibitem{wood2006}
B.~Wood, J.~B. Pendry, and D.~P. Tsai.
\newblock Directed subwavelength imaging using a layered metal-dielectric
  system.
\newblock {\em Phys. Rev. B}, 74:115116, 2006.

\bibitem{anantharamakrishna2003raa}
S.~A. Ramakrishna and J.~B. Pendry.
\newblock Removal of absorption and increase in resolution in a near-field lens
  via optical gain.
\newblock {\em Phys. Rev. B}, 67(20):201101, 2003.

\bibitem{ponizovskaya2007mni}
E.~V. Ponizovskaya and A.~M. Bratkovsky.
\newblock Metallic negative index nanostructures at optical frequencies: losses
  and effect of gain medium.
\newblock {\em Appl. Phys. A}, 87(2):161--165, 2007.

\bibitem{Saleh}
B.~Saleh and M.~Teich.
\newblock {\em Fundamentals of Photonics}.
\newblock John Wiley \& Sons, Inc, 1991.

\bibitem{GoodmanFourierOptics}
J.~W. Goodman.
\newblock {\em Introduction to Fourier Optics}.
\newblock McGraw-Hill, 1996.

\bibitem{dorofeenko2006fwa}
A.~V. Dorofeenko, A.~A. Lisyansky, A.~M. Merzlikin, and A.~P. Vinogradov.
\newblock Full-wave analysis of imaging by the pendry-ramakrishna stackable
  lens.
\newblock {\em Phys. Rev. B}, 73(23):235126, 2006.

\bibitem{nietovesperinas2004pis}
M.~Nieto-Vesperinas.
\newblock Problem of image superresolution with a negative-refractive-index
  slab.
\newblock {\em J. Opt. Soc. Am. A}, 21(4):491--498, 2004.

\bibitem{melville2006josab}
D.~O. Melville and R.~J. Blaikie.
\newblock Experimental comparison of resolution and pattern fidelity in single-
  and double-layer planar lens lithography.
\newblock {\em J. Opt. Soc. Am. B}, 23(3):461--467, 2006.

\bibitem{Scalora2007opex}
M.~Scalora, G.~D'Aguanno, N.~Mattiucci, M.~J. Bloemer, D.~Ceglia, M.~Centini,
  A.~Mandatori, C.~Sibilia, N.~Akozbek, M.~G. Cappeddu, M.~Fowler, and J.W.
  Haus.
\newblock Negative refraction and sub-wavelength focusing in the visible range
  using transparent metallo-dielectric stacks.
\newblock {\em Opt. Express}, 15:508--523, 2007.

\bibitem{Scalora1998ap}
M.~Scalora, M.~J. Bloemer, A.~Manka, S.~Pethel, J.~Dowling, and C.~Bowden.
\newblock Transparent, metallo-dielectric one dimensional photonic band gap
  structures.
\newblock {\em Appl. Phys.}, 83:2377, 1998.

\bibitem{belov2006prb}
P.~A. Belov and Y.~Hao.
\newblock Subwavelength imaging at optical frequencies using a transmission
  device formed by a periodic layered metal-dielectric structure operating in
  the canalization regime.
\newblock {\em Phys. Rev. B}, 73:113110, 2006.

\bibitem{belov2005csi}
P.~A. Belov, C.~Simovski, and P.~Ikonen.
\newblock Canalization of subwavelength images by electromagnetic crystals.
\newblock {\em Phys. Rev. B}, 71(19):193105, 2005.

\bibitem{ikonen2006eds}
P.~Ikonen, P.~A. Belov, C.~Simovski, and S.~Maslovski.
\newblock Experimental demonstration of subwavelength field channeling at
  microwave frequencies using a capacitively loaded wire medium.
\newblock {\em Phys. Rev. B}, 73(7):073102, 2006.

\bibitem{BornWolf}
M.~Born and E.~Wolf.
\newblock {\em Principles of Optics}.
\newblock Cambridge Univ. Press, 7th edition, 1999.

\bibitem{Taflove:FDTD}
A.~Taflove and S.~C. Hagness.
\newblock {\em Computational Electrodynamics: The Finite-Difference Time-Domain
  Method}.
\newblock Artec House Inc., Boston, second edition, 2000.

\bibitem{Farjadpour2006ol}
A.~Farjadpour, D.~Roundy, A.~Rodriguez, M.~Ibanescu, P.~Bermel, J.~D.
  Joannopoulos, S.~G. Johnson, and G.~Burr.
\newblock Improving accuracy by subpixel smoothing in fdtd.
\newblock {\em Opt. Lett.}, 31:2972--2974, 2006.

\bibitem{Kotynski2008}
R.~Kotynski, K.~Krol, J.~Pniewski, and K.~Panajotov.
\newblock Analysis of two-dimensional polarisation-coupled impulse response in
  multilayered metallic flat lens.
\newblock {\em Proc. SPIE}, 6987--15(15), 2008.

\bibitem{Kotynski2007oqe}
R.~Kotynski, M.~Dems, and K.~Panajotov.
\newblock Waveguiding losses of micro-structured fibres--plane wave method
  revisited.
\newblock {\em Opt. Quantum Electron.}, 39:469--479, 2007.

\bibitem{JohnsonChristy}
P.~B. Johnson and R.~W Christy.
\newblock Optical constants of the noble metals.
\newblock {\em Phys. Rev. B}, 6:4370--4379, 1972.

\bibitem{Bloemer1998apl}
M.~J. Bloemer and M.~Scalora.
\newblock Transmissive properties of {Ag/MgF2} photonic band gaps.
\newblock {\em Appl. Phys. Lett}, 72:1676, 1998.

\bibitem{zhao2007prb}
Yan Zhao, Pavel Belov, and Yang Hao.
\newblock Accurate modeling of the optical properties of left-handed media
  using a finite-difference time-domain method.
\newblock {\em Phys. Rev. B}, 75:037602, 2007.

\end{thebibliography}

\end{document}